\documentclass[aps,prx,reprint, amsmath, amssymb,superscriptaddress]{revtex4-1}

\usepackage{bm}
\usepackage[retainorgcmds]{IEEEtrantools}
\usepackage{graphicx}
\usepackage{mathrsfs}
\usepackage{amsmath}
\usepackage{amssymb}
\usepackage{color}
\usepackage{amsfonts}
\usepackage{times,txfonts}
\usepackage{nicefrac}
\usepackage[colorlinks=true,linkcolor=blue,urlcolor=blue,citecolor=blue,pdfusetitle]{hyperref}

\usepackage[dvipsnames]{xcolor}
\definecolor{mypink}{rgb}{0.858, 0.188, 0.478}
\definecolor{bondiblue}{rgb}{0.0, 0.58, 0.71}
\definecolor{bleudefrance}{rgb}{0.19, 0.55, 0.91}

\newcommand{\tr}{\text{tr}}
\newcommand{\trans}{^{\text{T}}}

\begin{document}

\title{Correlations break homogenization }
\date{\today}
\author{Naim Elias Comar}
    \affiliation{Instituto de F\'isica da Universidade de S\~ao Paulo,  05314-970 S\~ao Paulo, Brazil}
\author{Gabriel T. Landi}
    \affiliation{Instituto de F\'isica da Universidade de S\~ao Paulo,  05314-970 S\~ao Paulo, Brazil}

\begin{abstract}
The standard collisional model paradigm consists of a system that interacts sequentially with identically prepared ancillas.
After infinitely many collisions, and under appropriate conditions, the system may converge to the same state as the ancillas. 
This process, known as homogenization, is independent of the ancilla initial state, being a property only of the underlying dynamics. 
In this paper we extend this idea to locally identical, but globally correlated, ancillas, and show that correlations break homogenization.
This is done numerically using a minimal qubit model, and analytically using an exactly soluble  Gaussian model. 
In both cases, we use Hamiltonian graph states with cyclic graphs as the prototypical method for building scalable many-body entangled ancillary states.

\end{abstract}

\maketitle{}

%
%
\section{\label{sec:int}Introduction}
%
%

Relaxation towards equilibrium is one of the most basic problems in non-equilibrium physics. 
In classical thermodynamics, this process is often taken as a universal tendency, which every system naturally undergoes. 
This is deeply associated with Boltzmann's molecular chaos hypothesis (\emph{Stosszahlansatz}). 
For instance, when a hot particle enters a bath composed of cold particles, it will suffer multiple collisions, each involving the exchange of a certain amount of energy. 
This will lead, after  sufficiently many collisions, to the hot particle homogenizing with the cold bath. 
This is thus directly associated to the spreading of energy and information among multiple degrees of freedom.

As one moves to the nano or quantum domain, relaxation becomes more dependent on finer details about the environment, as well as the system-environment interactions.
Features such as strong coupling~\cite{Campisi2011,Perarnau-Llobet2018a,Talkner2020} and non-Markovianity~\cite{Breuer2015,Rivas2014}, all affect the relaxation in fundamental ways, making this an active field of research. 
A particularly interesting scenario is that of \emph{homogenization}, first put forth in~\cite{Ziman2002,Scarani2002}. 
The authors considered a qubit system interacting with infinitely many independent and identically prepared (iid) qubit ancillas.
They then studied under which conditions the system would tend, in the limit of infinitely many collisions, to the same state as the ancillas. 
This is, in a sense, a generalization of thermalization, because it is independent of the state of the ancillas; instead, it depends only on the properties of the dynamics. 
In fact, the authors showed that, for qubits, the unitary achieving this task was the partial swap.

The scenario in Refs.~\cite{Ziman2002,Scarani2002} is an instance of the now-popular Collisional Models (CMs)~\cite{Rau1963,Englert2002,Strasberg2016,DeChiara2018,Landi2020a}. 
CMs model open system dynamics in a stroboscopic fashion, via sequential collisions with different ancillas. 
They are interesting for a variety of reasons. 
First, they allow full control over which ingredients  are introduced. 
This may include ancilla-ancilla correlations, which lead to non-Markovianity\cite{Campbell2019a,Man2018,Lorenzo2017,Donvil2021,Mascarenhas2017,McCloskey2014,Campbell2018b,Rybar2012a,Cilluffo,Ciccarello2013b,Bernardes2017,Taranto2018a,Ciccarello2013a,Filippov2017,Kretschmer2016,Jin2018,Ciccarello2013,Bernardes2014,Cakmak2017,Daryanoosh2018,DeChiara2020}, work-driven unitaries, which are used to implement heat engines~\cite{Quan2007,Allahverdyan2010,Uzdin2014,Campisi2014,Campisi2015,Denzler2019,Pezzutto2019,Mohammady2019,Molitor2020} and continuous measurements in the ancillas to implement stochastic unravellings~\cite{Gross2017a,Rossi2020,Landi2021a}.
Second, CMs allow for great control over the thermodynamics of quantum systems, and have been used to clarify a series of issues about the validity of the first and second laws~\cite{Barra2015,Pereira2018,DeChiara2018}.
Finally, the dynamics of CMs only involve a few degrees of freedom at a time, making it more manageable than standard open quantum system treatments, where the bath is macroscopic. 

\begin{figure*}
    \centering
    \includegraphics[width=0.8\textwidth]{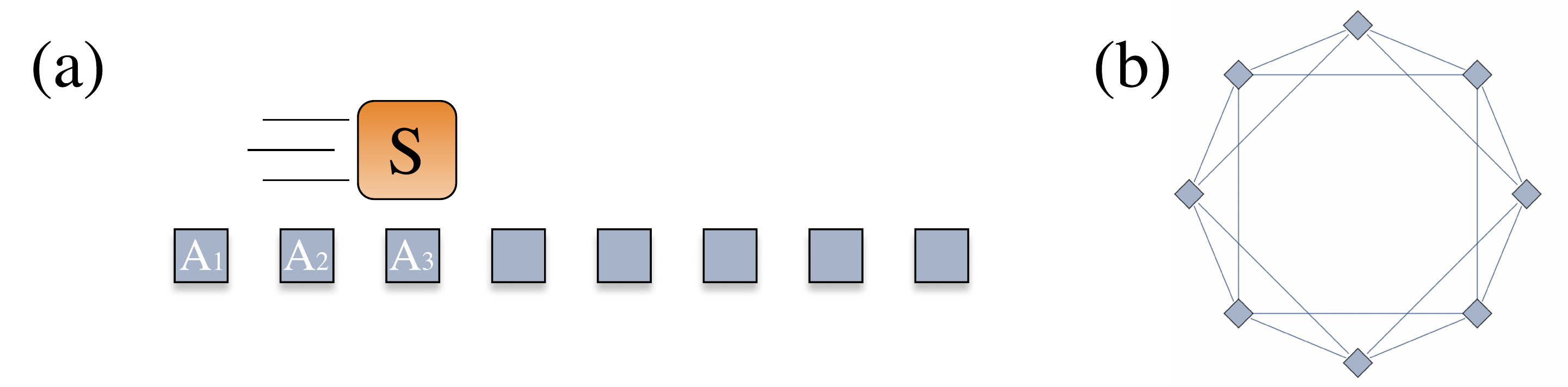}
    \caption{(a) Schematics of a collisional model, where the system $S$ interacts sequentially with a series of ancillas $A_1,A_2,\ldots$.
    (b) The ancillas are not initially independent, however. Instead, they are in a correlated state $\rho_{\bm A}$, which we choose here to construct as a Hamiltonian cyclic graph state. In this example, we have a cyclic graph of the form~\eqref{G}, with $c_{1,2}\neq 0$ and $c_i=0$ for $i>2$.
    }
    \label{fig:drawing}
\end{figure*}

Notwithstanding, homogenization still remains a widely unexplored: \emph{after infinitely many collisions, to which state will the system eventually relax to?}
The goal of this paper is to show that homogenization is broken when the ancillas are locally identical, but globally correlated. 
The idea is shown in Fig.~\ref{fig:drawing}(a): a system $S$, prepared in a generic state $\rho_S^0$, is put to interact with infinitely many ancillas $A_1, A_2, \ldots$. Locally, they are all prepared in the same state $\rho_{A_i} \equiv \rho_{A}$. But globally, they are in a correlated state $\rho_{A_1A_2,\ldots}\equiv \rho_{\bm A}$. 
As we show, in the long-time limit the steady-state of the system will differ from the ancilla's local state, by an amount which depends on the structure of the correlations. 
This is interesting because, as far as the system is concerned, it is always interacting locally with identically prepared ancillas. 
But globally, the correlations play a non-trivial role,  steering the system away from the homogenized state. 

To draw an analogy, suppose PhD students want to convince their supervisor to buy a new coffee machine.
To accomplish that, they go to the supervisor's office, one at a time, and lay down their arguments as to why a new machine is absolutely necessary. 
After very many ``collisions'', the supervisor will eventually make up their mind whether to buy it or not.
This would be the standard homogenization approach.
Instead, suppose  that before interacting with their supervisor, the students meet up in order to align their speeches. They may share ideas on what are the most persuasive arguments and how to best convince the supervisor. 
After establishing this pre-defined narrative, they then go to the supervisor's office, again one at a time. 
Clearly, whether they align their speeches or not should have a dramatic effect on the outcome of the process and, hopefully, should help them get the new coffee machine.

The scenario just described is an example of a common-cause memory, and has been widely used in the study of non-Markovianity~\cite{Man2018,Mascarenhas2017,Rybar2012a,Filippov2017,Bernardes2014,Cakmak2017}.
This is to be constrasted with direct cause CMs~\cite{Camasca2020,Ciccarello2013a,McCloskey2014,Kretschmer2016,Ciccarello2013,Cakmak2017,Seah2019,Shu2020}, where the correlations between ancillas are introduced midway through the dynamics. 
Our interest here is different, however. 
Instead of analyzing the information backflow, as is usually done, we study the final state of the system, and how this is affected by the correlations.

In Sec.~\ref{sec:idea} we lay down the main idea behind homogenization, and how we propose to study its violations using correlated ancillas. 
Then, in Sec.~\ref{sec:qubit} we analyze a minimal qubit model, which has to be computed numerically, but nonetheless clearly illustrates the effect. 
Finally, in Sec.~\ref{sec:CV}, we move to continuous variable Gaussian states, where analytical results can be provided.


\section{\label{sec:idea}Formal Framework}

The Hilbert space of $S$ is chosen to be isomorphic to that of $A_i$. 
We let $U_i$ denote a local interaction between S and $A_i$, and choose $U_i$ to be homogenizing in the sense of~\cite{Ziman2002,Scarani2002}. 
This means that
\begin{equation}\label{partial_swap}
    U_i (\rho_{A_i} \otimes \rho_{A_i}) U_i^\dagger = \rho_{A_i} \otimes \rho_{A_i}.
\end{equation}
For qubits, the class of unitaries satisfying this are partial swaps.
This kind of property has also been widely used in resource theories.
If $\rho_{A_i} = \mathbb{I}/d$ is a maximally mixed state (where $d$ is the Hilbert space dimension of each ancilla), this becomes akin to the resource theory of purity~\cite{Horodecki2003}. 
And if $\rho_{A_i}$ is a thermal state, Eq.~\eqref{partial_swap} becomes a thermal operation~\cite{Janzing2000,Horodecki2013,Brandao2013}. 
For our purposes, the states $\rho_{A_i}$ are generic. 

{\color{black}The choice of unitary~\eqref{partial_swap} can be further motivated as follows. 
First, provided one is interested only in local unitaries, the partial swap is meaningful in itself, as an operation that partially exchanges the quantum states of the two systems.
This, together with the fact that the ancilla's local states are all identical, make the problem as close as possible to the original homogenization scenario.
Second, the local nature of the unitary also makes it so that any local observer, with access only to the reduced state of the system, is unable to ascertain that the system is not homogenizing. This would only possible \emph{a posterior}, by comparing the final state of the system with those of the ancillas. 

}

Each ancilla participates only once in the dynamics. 
Hence, the global state of $S,A_1,A_2,\ldots$ after $n$ collisions will be given by 
\begin{equation}\label{global_map}
    \rho_{S{\bm A}}^{n} = U_n \ldots U_1 (\rho_S^0 \otimes \rho_{\bm A} ) U_1^\dagger \ldots U_n^\dagger.
\end{equation}
Since the ancillas are initially correlated, even the very first collision (with $A_1$) will already cause $S$ to become correlated with all other ancillas $A_2,A_3,\ldots$. 
This is the origin of non-Markovianity and implies that it is impossible to break the map~\eqref{global_map} in terms of smaller maps from $n$ to $n+1$. 
We focus on the reduced state of the system, $\rho_S^n = \tr_{\bm A}~ \rho_{S {\bm A}}^n$. 
And, in particular, in the long-time fixed point 
\begin{equation}
    \rho_S^* = \lim\limits_{n\to\infty}~\rho_S^n. \label{fixed_point}
\end{equation}
Our goal is to investigate how correlations in $\rho_{\bm A}$ affect $\rho_S^*$. 

One of the challenges in formalizing the above idea is the choice of initial ancilla state $\rho_{\bm A}$. 
To make the problem tractable, we postulate that the reduced states should be identical: $\rho_{A_i} = \tr_{{\bm A}/i}~\rho_{\bm A} = \rho_{A_1}$, where $\tr_{{\bm A}/i}$ means the partial trace over all ancillas, except the $i$-th. 
This implies that, as far as local operations are concerned, the system is just interacting with identical ancillas.

In addition, we also wish for these states to be built in a scalable way, so that we can consider arbitrarily many ancillas. 
After all, homogenization refers to infinitely many collisions. 
We have therefore found it convenient to choose $\rho_{\bm A}$ to be \emph{translationally invariant}. 
That is, for any group $\{A_k,A_{k+1},\ldots,A_{k+\ell}\}$, we impose that the reduced state 
\begin{equation}
    \rho_{A_k A_{k+1}\ldots A_{k+\ell}} = \tr_{{\bm A}/\{k,\ldots,k+\ell\}} ~\rho_{\bm A} = \rho_{A_1\ldots A_{1+\ell}},
\end{equation}
should be independent of $k$; i.e., a function only of the block length $\ell$.

In this paper we implement this using Hamiltonian cyclic graph states~\cite{Hein2004,Pantaleoni2021,Aolita2011,Gu2009a,Menicucci2011} (Fig.~\ref{fig:drawing}(b)). 
We consider a system with a finite number $N_A$ of  ancillas, and in the end, take $N_A \to \infty$. 
Let $H_{ij}$ denote a certain Hamiltonian interaction  between ancillas $i$ and $j$. 
This may be, e.g. $\sigma_x^i \sigma_x^j$ for qubit systems, where $\sigma_x^i$ are Pauli matrices. 
The global initial state $\rho_{\bm A}$ is chosen to be of the form $\rho_{\bm A} = |\Psi_{\bm A}\rangle\langle \Psi_{\bm A}|$, where 
\begin{equation}\label{psi_A}
    |\Psi_{\bm A}\rangle = e^{- i k\sum_{i,j} G_{ij} H_{ij}} |\phi\rangle^{\otimes N_A}.
\end{equation}
Here $|\phi\rangle$ are arbitrary single ancilla states, $k$ is an interaction strength and the $G$ (of size $N_A\times N_A$) is the graph adjacency matrix; that is, the coefficients $G_{ij}$ represent the magnitude with which ancilla $i$ couples to ancilla $j$.
What makes $\rho_{\bm A}$ translationally invariant is the choice for  $G$. 
In particular, we choose it as a \emph{symmetric circulant} matrix. 
For, e.g., $N_A = 5$, such a matrix has the form
\begin{equation}\label{G}
    G = \begin{pmatrix}
    0   & c_1 & c_2 & c_3 & c_2 & c_1 \\
    c_1 & 0   & c_1 & c_2 & c_3 & c_2 \\
    c_2 & c_1 & 0   & c_1 & c_2 & c_3 \\
    c_3 & c_2 & c_1 & 0   & c_1 & c_2 \\
    c_2 & c_3 & c_2 & c_1 & 0   & c_1 \\
    c_1 & c_2 & c_3 & c_2 & c_1 & 0 
    \end{pmatrix},
\end{equation}
for arbitrary real coefficients $c_1$. 
This is the adjacency matrix for a cyclic graph, as illustrated in  Fig.~\ref{fig:drawing}(b).
For instance, if $c_1 \neq 0$ and all other $c_i = 0$, the graph represents a nearest-neighbor coupling.
But this does not mean that the correlations will be only between nearest neighbors, due to the exponential in~\eqref{psi_A}. 
This method is not the most general way of constructing translationally invariant states. 
But it provides a systematic way of doing so for arbitrary system sizes $N_A$. 
We also stress that the preparation~\eqref{psi_A} is done \emph{before} the actual dynamics~\eqref{global_map}.

%
%
\section{\label{sec:qubit}Minimal qubit model}
%
%

In order to illustrate the main idea, we  first consider a numerical example where system and ancillas are composed of qubits.
The system is assumed, for simplicity, to start in the computational basis state $|0\rangle$. 
To comply with~\eqref{partial_swap}, we assume that the they interact via a partial swap unitary~\cite{Ziman2002,Scarani2002} 
\begin{equation}\label{exchange_interaction}
U_n = \exp\{ - i \tau (\sigma_S^+ \sigma_{A_n}^- + \sigma_S^- \sigma_{A_n}^+)\},    
\end{equation}
where $\tau$ controls the interaction strength.
{\color{black}The case  $\tau = \pi/2$ represents the full swap, $U|\phi, \psi\rangle = |\psi,\phi\rangle$. This case is special because it always pushes the system towards the exact reduced state of the ancillas (which are all equal). 
For instance,  we suppose just for now that the system is in a pure state $|\phi\rangle$, and the ancillas are in a correlated state $|\psi_{A_1A_2A_3\ldots}\rangle$.
Then, after applying say $U_1$, we would get 
\begin{equation}
    U_1 \Big(|\phi\rangle_S \otimes |\psi\rangle_{A_1A_2A_3\ldots}\Big) = |\psi\rangle_{SA_2A_3\ldots} \otimes |\phi\rangle_{A_1}.
\end{equation}
Hence, due to the full swap the system does become correlated with the ancillas. But locally, it is always pushed towards the local state of the ancillas. The same logic applies to the action of $U_2$, $U_3$, and so on. 
For this reason, we will henceforth be particularly interested in partial swaps, for which the bahvior is richer. 
}  

\subsection{Preparation of the initial ancilla state}

\begin{figure*}
    \centering
    \includegraphics[width=0.95\textwidth]{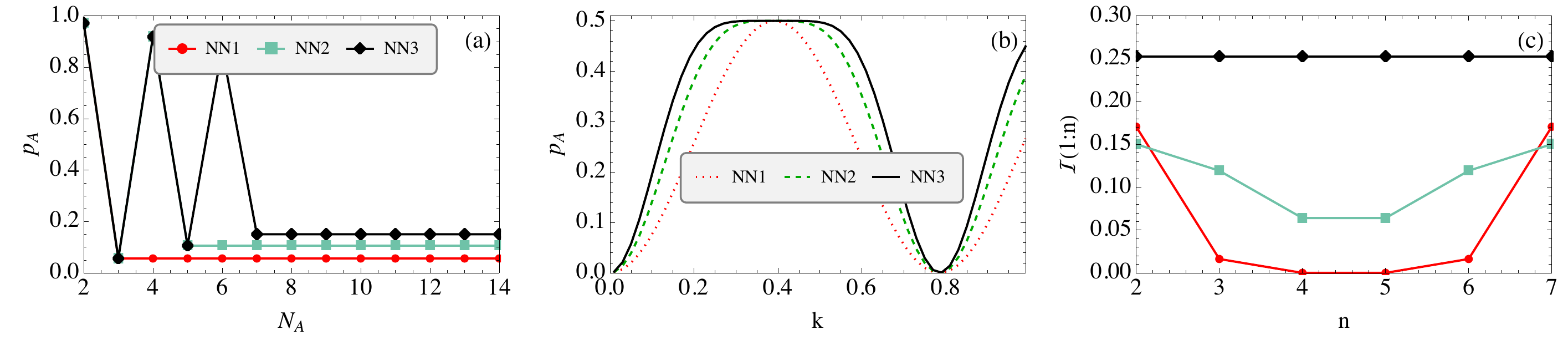}
    \caption{Properties of the initial ancilla state $\rho_{\bm A}$ for different choices of graph state nearest-neighbor interactions, NN1, NN2 and, NN3.
    (a) Excited state populations $p_A=\langle 1 | \rho_{A_1}|1\rangle$ of the reduced ancilla states $\rho_{A_i} \equiv \rho_{A_1}$ for varying graph sizes $N_A$, with fixed $k = 0.7$. 
    When $N_A \to \infty$ the population no longer changes with $N_A$.  
    (b) $p_A$ as a function of the overall strength $k$ [Eq.~\eqref{psi_A}] for fixed $N_A = 7$ ancillas.
    (c) Mutual information $\mathcal{I}(1{:}n)$ between the first and the $n$-th ancillas, as a function of $n$, with fixed $N_A = 7$ and $k = 0.7$. 
    For nearest-neighbors larger than 3, the mutual information becomes flat, meaning all ancillas are almost equally correlated.
    }
    \label{fig:qubit_ancilla}
\end{figure*}

We prepare the initial ancilla state~\eqref{psi_A}
assuming $|\phi\rangle = |0\rangle$ and $H_{ij} = \sigma_x^i \sigma_x^j$. 
The connections are in the form of the circulant graph~\eqref{G}.
We always take all $c_i$ to be either 0 or 1 and fix the overall strength at $k = 0.7$, except when stated otherwise. 
We then analyze what happens when the interactions are only first nearest-neighbors ($c_1=1$ and $c_i = 0$ for $i >1$), second nearest neighbors, $c_1=c_2 = 1$ and $c_i=0$ for $i > 2$) and third nearest-neighbors. 
We refer to these as NN1, NN2 and NN3 respectively. 
Fig.~\ref{fig:qubit_ancilla} illustrates how the initial state $\rho_{\bm A}$ depends on the properties of the graph states. 
Locally, the states of the ancillas are diagonal in the computational basis.  
This can be lifted by considering other choices of $|\phi\rangle$, for instance. 

Fig.~\ref{fig:qubit_ancilla}(a) shows, for fixed $k= 0.7$, that the excited state populations of the local states $p_A=\langle 1 | \rho_{A}|1\rangle$ become eventually independent of $N_A$ as it increases. 
Since the state is translationally invariant, this holds true for all local states of the ancillas, $\rho_{A_i} \equiv \rho_{A}$. 

In Fig.~\ref{fig:qubit_ancilla}(b) we show how $p_A$ varies in terms of the overall interaction strength $k$ in Eq.~\eqref{psi_A}. 
Since for qubits a diagonal state is tantamount to a thermal state, one can view $p_A$ as representing an effective local temperature of the ancillas. 
As $k$ is varied, one can go from a situation where $p_A \sim 0$ all the way to $p_A \sim 1/2$, which stands for a maximally mixed state (infinite temperature). 
The local states of the ancillas may thus be tuned arbitrarily by varying the correlation strength $k$. 

Finally, Fig.~\ref{fig:qubit_ancilla}(c) shows how the correlations in the global state $\rho_{\bm A}$ behave. 
We analyze here the mutual information between ancillas 1 and $n = 2,3,\ldots, N_A$, which is given by 
\begin{equation}
   \mathcal{I}(1:n) = S(\rho_{A_1}) + S(\rho_{A_n}) - S(\rho_{A_1,A_n}), 
\end{equation}
with $S(\rho) = - \tr (\rho \ln \rho)$ being the von Neumann entropy. 
As can be seen in the image, the correlations first decay with $n$ and then grow again. This is due to the cyclic nature of the graph. 
However, significant variations are only observed for first nearest-neighbors (NN1; red circles). 
As one moves to NN2 and NN3, the correlations quickly become less dependent on $d$. 
In particular, for NN3 the correlation profile is practically flat, meaning that all ancillas are roughly equally correlated. 

\begin{figure*}
    \centering
    \includegraphics[width=0.85\textwidth]{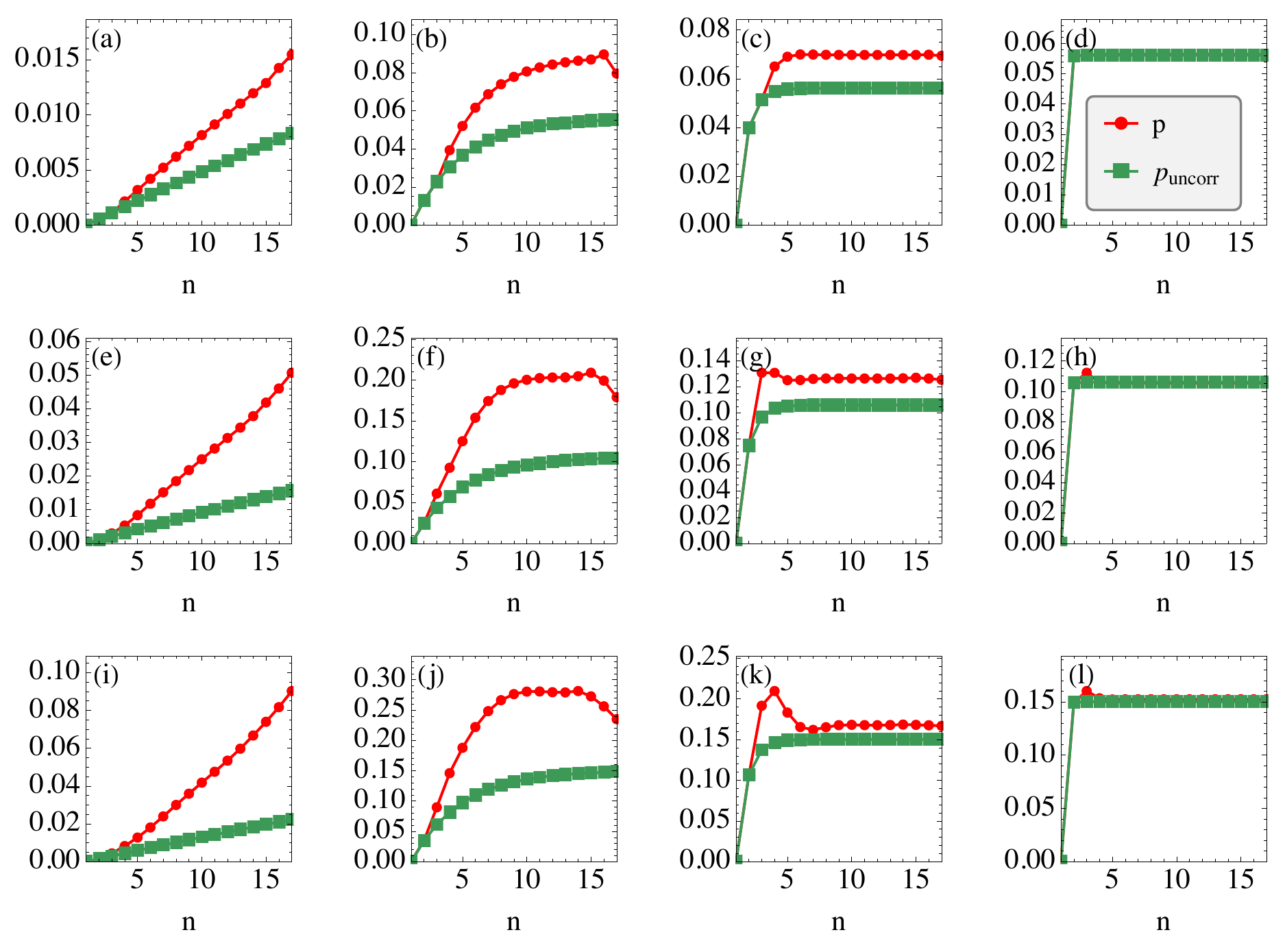}
    \caption{Dynamics of the system excited state population $p$ as a function of the collision number $n$. 
    Each line corresponds to NN1, NN2 and NN3, while each row corresponds to $\tau = 0.1, 0.5, 1.0, 1.5$ [Eq.~\eqref{exchange_interaction}]. 
    We fixed $k = 0.7$. 
    Green squares correspond to the dynamics with the ancillas in the same local state $\rho_{A_i}$, but completely uncorrelated.
    All curves were computed with $N_A = 16$ ancillas.
    }
    \label{fig:qubit}
\end{figure*}

\subsection{System dynamics}

Having analyzed the structure of $\rho_{\bm A}$, we now turn to how this affects the dynamics of the system. 
Fig.~\ref{fig:qubit} plots the system's excited state population $p = \langle 1| \rho_S | 1\rangle$ as a function of the collision number $n$ (stroboscopic time). 
Each row corresponds to different choices of graph states, NN1, NN2 and NN3. 
Each column refers to a different interaction strength $\tau$ [Eq.~\eqref{exchange_interaction}]. 
All curves were computed with $k = 0.7$ and $N_A = 16$ ancillas, which was the maximum number of ancillas we were able to reach.
For reference, we compare this  with the dynamics that would result from using independent ancillas, with the same reduced state $\rho_{A_1}$. 
That is, a reference process where the initial state is instead $\rho_{\bm A}^{\rm uncorr} = \rho_{A_1} \otimes \rho_{A_2} \otimes \ldots$.
The resulting probabilities are referred to as $p_{\rm uncorr}$.

When $\tau$ is small (first column), the transfer of information in each collision is small, causing the evolution of $p$ to be slower. 
The steady-state value is not clear in this case, however: 
Since we are only able to reach $N_A = 16$ ancillas, it is not possible to known to which state $p$ will eventually converge. 
This is one of the main difficulties in dealing with qubit systems. 
Conversely, if $\tau = 1.5$ (fourth column of Fig.~\ref{fig:qubit}), the information transfer is very large and the system quickly homogenizes. However, the effects of the correlations are now small and the final population $p$ is very close to $p_{\rm uncorr}$.  

The interesting situations occur for intermediate $\tau$. As can be seen, the dynamics of $p$ and $p_{\rm uncorr}$ clearly split from each other and, in the long time limit, converge to different values. 
This shows how correlations break homogenization. 
In these examples we always have $p> p_{\rm uncorr}$, meaning that the final state is hotter than it would be if correlations were not present. 
But it does not have to be this way and, using different values of $k$ and $\tau$ it is possible to also reach situations where $p < p_{\rm uncorr}$.

{\color{black}
\subsection{Dynamics of the ancilla-ancilla mutual information}

Fig.~\ref{fig:qubit_ancilla}(c) shows the mutual information $\mathcal{I}(1{:}n)$ between the first and $n$-th ancilla, in the prepared graph state $|\psi_{\bm A}\rangle$. 
As the ancillas start to collide with the system, however, this correlation profile will start to change. 
In Ref.~\cite{Ziman2002} the authors studied how the correlations among the ancillas develop in time, due to their collisions with the system. 
Here we perform a similar analysis to contrast with those results. We focus instead on the mutual information, as in Fig.~\ref{fig:qubit_ancilla}(c). 

Fig.~\ref{fig:MI_dynamics} presents the dynamics of $\mathcal{I}(1{:}n)$ after each collision. 
Image (a) concerns the Markovian case, while (b) and (c) correspond to first and third nearest-neighbors. 
The Markovian case starts with zero correlations (by default) and produces an exponentially decaying profile $I(1{:}n) \sim e^{-\alpha k}$, similar to what was found in Ref.~\cite{Ziman2002}, although there the authors focused on the concurrence (we prefer the mutual information here because the initial ancilla state is mixed). 
Conversely, when correlations are present, the profile is already non-zero before the collisions. As the dynamics evolves, this correlation profile is then distorted. 
Due to the finite number of ancillas we can simulate, we are not able to observe any emerging patterns. Overall, the correlations with different ancillas can both decrease or increase, in non-trivial ways. 
This is particularly visible in the lower panels of Fig.~\ref{fig:MI_dynamics}, which plots the evolution of the mutual information in time.

\begin{figure*}
    \centering
    \includegraphics[width=\textwidth]{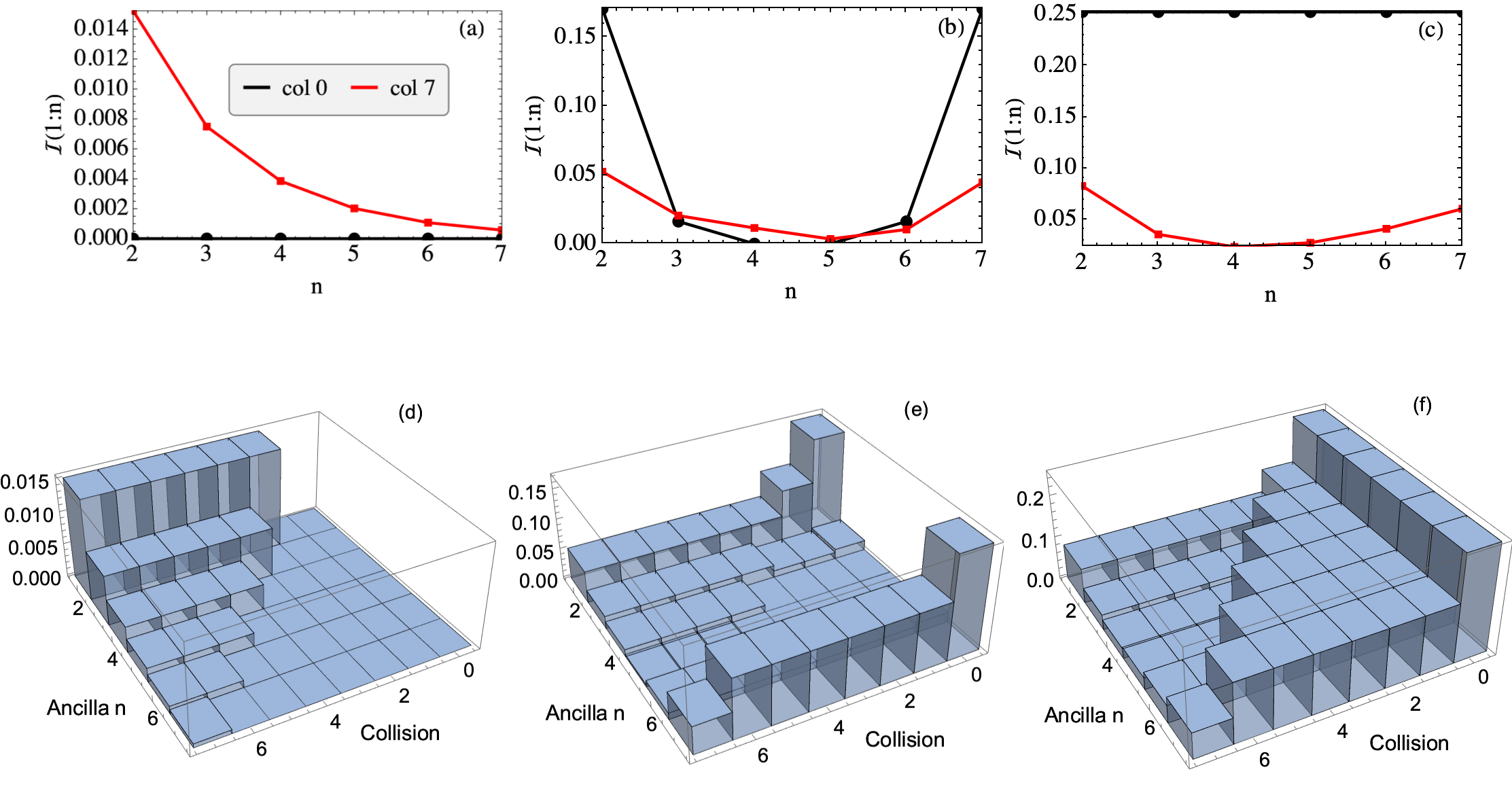}
    \caption{Dynamics of the mutual information between the first and $n$- ancillas.
    Upper panel: before and after all collisions. 
    Lower panel: at each collision step.
    (a),(d) For the Markovian case. 
    (b),(e) Non-Markovian case with nearest-neighbor interactions. 
    (c),(f) Same, but with 3 nearest-neighbors. 
    In (b) and (c), the profiles of $\mathcal{I}(1{:}n)$ before the first collision (black) are the same as those in Fig.~\ref{fig:qubit_ancilla}(c).
    Parameters: $k = 0.7$ and $\tau = 1.0$.
    }
    \label{fig:MI_dynamics}
\end{figure*}

}

%
%
\section{\label{sec:CV}Continuous variable Gaussian models}
%
%

In order to obtain deeper insights into how correlations affect homogenization, we now turn to a class of exactly soluble models with arbitrarily many ancillas. 
Namely, we assume that both system and ancillas are given by bosonic modes, with quadratures $(q_S,p_S)$ and $(q_i, p_i)$, satisfying canonical commutation relations.
The system-ancilla interactions are assumed to be given by a beam-splitter unitary 
\begin{equation}\label{beam_splitter}
    U_n = e^{\tau (a_S^\dagger a_i - a_i^\dagger a_S)},
\end{equation}
where $a_\alpha = (q_\alpha + i p_\alpha)/\sqrt{2}$, $\alpha = S,1,2,\ldots,N_A$. 
This is the bosonic analog of the partial swap~\eqref{partial_swap}. 

The unitary~\eqref{beam_splitter} is Gaussian preserving. 
We assume that initially system and ancilla are all in Gaussian states. Hence, they will remain so throughout.
The entire dynamics can thus be conveniently described in terms of the system covariance matrix (CM), which is given by 
\begin{equation}
    \gamma_{\alpha,\beta} = \frac{1}{2} \langle \{ R_\alpha, R_\beta\} \rangle - \langle R_\alpha \rangle\langle R_\beta\rangle,
\end{equation}
where $R = (q_S,p_S,q_1,p_1,\ldots,q_{N_A},p_{N_A})$ is a vector of length $2N_A +2$.
We assume, without loss of generality, that the first moments $\langle R_\alpha\rangle$ are initially zero.
Due to the beam-splitter unitary~\eqref{beam_splitter}, they then remain so throughout. 

Initially, since system and ancilla are uncorrelated, the global state will be of the form
(e.g. for $N_A = 5$)
\begin{equation}
    \gamma_{S {\bm A}}^0 = 
    \begin{pmatrix}
    \gamma_S^0 & 0  & 0 & 0 & 0 & 0 \\[0.2cm]
   0 & \gamma_{A_1} & \zeta_{1,2} & \zeta_{1,3} & \zeta_{1,4} & \zeta_{1,5}  \\[0.2cm]
    0  & \zeta_{1,2}\trans & \gamma_{A_2} & \zeta_{2,3} & \zeta_{2,4} & \zeta_{2,5}  \\[0.2cm]
    0  & \zeta_{1,3}\trans & \zeta_{2,3}\trans & \gamma_{A_3} & \zeta_{3,4} & \zeta_{3,5}  \\[0.2cm]
    0  & \zeta_{1,4}\trans & \zeta_{2,4}\trans & \zeta_{3,4}\trans & \gamma_{A_4} & \zeta_{4,5}  \\[0.2cm]
    0  & \zeta_{1,5}\trans & \zeta_{2,5}\trans & \zeta_{3,5}\trans & \zeta_{4,5}\trans & \gamma_{A_5} 
    \end{pmatrix},
\label{CV_initial_state}    
\end{equation}
where all elements are $2\times 2$ blocks.
For instance, 
\begin{equation}
    \gamma_S^0 = \begin{pmatrix}
    \langle q_S^2 \rangle & \frac{1}{2} \langle\{ q_S,p_S\}\rangle \\[0.2cm]
    \frac{1}{2} \langle\{ q_S,p_S\}\rangle & \langle p_S^2\rangle
    \end{pmatrix},
\end{equation}
is the initial CM of the system. 
Similarly $\gamma_{A_i}$ in Eq.~\eqref{CV_initial_state} are the reduced CMs of each ancilla, which are all equal, $\gamma_{A_i} = \gamma_A$. 
Moreover, $\zeta_{i,j}$ represent the correlations between $i$ and $j$. 
Since the states we choose are always translationally invariant, they have the convenient property that $\zeta_{j,j+d} = \zeta_{d}$; i.e., they only depend on the distance between the ancillas. 
The initial ancilla state will thus be of the form 
\begin{equation}
    \gamma_{\bm A} = \begin{pmatrix}
     \gamma_{A} & \zeta_{1} & \zeta_{2} & \zeta_{3} & \zeta_{4}  \\[0.2cm]
    \zeta_{1}\trans & \gamma_{A} & \zeta_{1} & \zeta_{2} & \zeta_{3}  \\[0.2cm]
    \zeta_{2}\trans & \zeta_{1}\trans & \gamma_{A} & \zeta_{1} & \zeta_{2}  \\[0.2cm]
    \zeta_{3}\trans & \zeta_{2}\trans & \zeta_{1}\trans & \gamma_{A} & \zeta_{1}  \\[0.2cm]
    \zeta_{4}\trans & \zeta_{3}\trans & \zeta_{3}\trans & \zeta_{1}\trans & \gamma_{A} 
    \end{pmatrix},
\label{CV_gamma_A}
\end{equation}
which is a Toeplitz matrix.

In terms of the CM, the unitary~\eqref{beam_splitter} is now replaced by multiplication by a symplectic matrix $S_n$, so that the global dynamics becomes
\begin{equation}
    \gamma_{S {\bm A}}^n = S_n~\gamma_{S {\bm A}}^{n-1}~S_n\trans. \label{sympletic_evolution}
\end{equation}
The matrix $S_n$ can be conveniently parametrized in $2\times 2$ blocks,  as 
(\textit{e.g.} for $N_A=5$)
\begin{align}
   & S_1= \begin{pmatrix}
      c & s & 0 & 0 & 0 & 0 \\
      -s & c & 0 & 0 & 0 & 0 \\
      0 & 0 & 1 & 0 & 0 & 0 \\
      0 & 0 & 0 & 1 & 0 & 0 \\
      0 & 0 & 0 & 0 & 1 & 0 \\
      0 & 0 & 0 & 0 & 0 & 1
     \end{pmatrix},
     ~ 
     S_2= \begin{pmatrix}
      c & 0 & s & 0 & 0 & 0 \\
      0 & 1 & 0 & 0 & 0 & 0 \\
      -s & 0 & c & 0 & 0 & 0 \\
      0 & 0 & 0 & 1 & 0 & 0 \\
      0 & 0 & 0 & 0 & 1 & 0 \\
      0 & 0 & 0 & 0 & 0 & 1
     \end{pmatrix}, ~ S_3= \cdots, \nonumber
\end{align}
where $c=\cos(\tau)$, $s=\sin(\tau)$. 
Here all entries are $2\times 2$ blocks, so $c$ actually means $c\mathbb{I}_2$, etc (where $\mathbb{I}_2$ is the $2\times 2$ identity matrix). 
For convenience, we will henceforth assume that $\zeta_{i,j}\trans = \zeta_{i,j}$. 
If this is not the case, one may simply replace $\zeta_{i,j} \to (\zeta_{i,j} +\zeta_{i,j}\trans)/2$ in all formulas below. This is possible because of the choice of phase in~\eqref{beam_splitter}, which treats all entries within each $2\times 2$ block in equal footing.

\subsection{General expression for the dynamics}

The convenient thing about the Gaussian model is that we can write down an explicit formula for the system covariance matrix at any time $n$. 
This can simply be done by inspection: we  write down $\gamma_S^n$ for the first few $n$ and construct the final result by induction. 
To a great extent, this is possible due to the very simple nature of the beam-splitter interaction~\eqref{beam_splitter}.
Another advantage is that this can be done without having to specify the actual initial ancilla state, which will be discussed in Sec.~\ref{sec:gaussian_initial_states}.

In any case, the system CM at time $n$, assuming generic elements $\gamma_{A_i}$ and $\zeta_{i,j}$ is
\begin{equation}
     \gamma_S^n = c^{2n}\gamma_S^0 + \sum_{j=1}^n c^{2(n-j)}s^2 \gamma_{A_j} + 2s^2\sum_{j=1}^{n-1}  \sum_{\ell>j}^n c^{2n-j-\ell}\zeta_{j,\ell}. \label{general_solution}
 \end{equation}
 Specializing to $\gamma_{A_i} = \gamma_A$ and $\zeta_{j,\ell} = \zeta_{|i-\ell|}$, allows us to  carry the first summation, simplifying the result to
 \begin{equation}
     \gamma_S^n = c^{2n}\gamma_S^0 + (1-c^{2n})\gamma_A + 
    2s^2\sum_{m=1}^{n-1}c^{2m} \sum_{d=1}^{m}c^{-d}\zeta_d. 
     \label{result_distance_dependence}
 \end{equation}
Recall that $\zeta_1$ is the nearest-neighbor correlation, $\zeta_2$ the second nearest-neighbor and so on, as in Eq.~\eqref{CV_gamma_A}.
If the ancillas are initially uncorrelated then $\zeta_d = 0$, and $\gamma_S^n$ will be a mixture of $\gamma_S^0$ and $\gamma_A$, with weights $c^{2n}$ and $1-c^{2n}$.
We assume that $c < 1$, so that when $n \to \infty$ we will always have $\gamma_S^* \to \gamma_A$. 
Hence, when correlations are absent the system always homogenizes.


\subsection{\label{sec:CV_particular}Particular cases}

It is insightful to consider some particular choices for the correlations $\zeta_d$.

\paragraph{Nearest-neighbors.} First suppose that $\zeta_1 = \zeta$ and $\zeta_d = 0$ for $d > 1$. 
Carrying out the sums in Eq.~\eqref{result_distance_dependence} results in
\begin{equation}
     \gamma_{S}^n = c^{2n}\gamma_S^0 + (1-c^{2n})\gamma_A + 2c(1-c^{2(n-1)})\zeta. \label{result_firstneighbor}
 \end{equation}
 Taking the limit $n \to \infty$ then leads to 
\begin{equation}
     \gamma_S^{*} = \gamma_A + 2c \zeta. \label{result_firstneighbor_ss}
\end{equation}
This beautifully illustrates how homogenization is broken by the correlation matrix $\zeta$. 
By changing the entries of $\zeta$, one may independently steer the entries of $\gamma_S^*$ away from those of $\gamma_A$. 
This result assumes $c \neq 0$ and $c \neq 1$. Otherwise, Eq.~\eqref{result_firstneighbor} would never actually reach a limiting value. 
Interestingly, though, the second term in~\eqref{result_firstneighbor} is proportional to $c$, meaning that the effects of the correlations will  only be absent if $\tau = \pi/2$, which corresponds to a full swap. 
{\color{black}This happens because the full swap forces the state of the system to always be in the state local reduced state of the ancillas, which are all equal by hypothesis.
In contrast, the effects of correlations }
are maximal in the case of extremely weak interactions $c\to 1$. 

\begin{figure}
    \centering
    \includegraphics[width=0.4\textwidth]{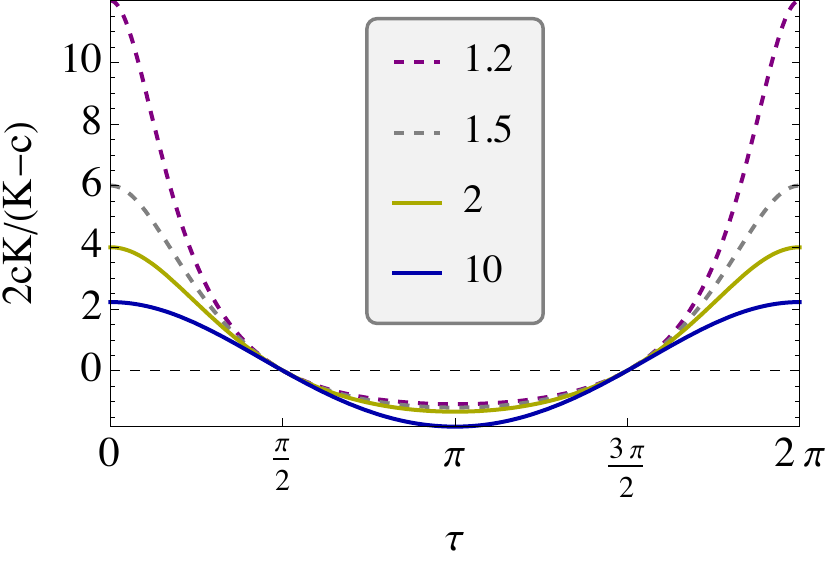}
    \caption{$2cK/K-c$~vs.~$\tau$, where $c=\cos(\tau)$,  for different values of $K > 1$. 
    Smaller $K$ represent longer-ranged interactions, for which the influence of correlations in the steady-state become more significant [Eq.~\eqref{result_exp_ss}]. 
    }
    \label{fig:correlations_influence_ss}
\end{figure}

\paragraph{Algebraically decaying correlations.}
As another interesting example, suppose that the correlations decay with the distance as
\begin{equation}
    \zeta_d = K^{1-d} \zeta, \qquad d = 1,2,\ldots,
    \label{correlations_exp}
\end{equation}
for some matrix $\zeta$ and a constant $K>1$. 
Longer-ranged interactions are represented by $K \gtrsim 1$, while short-range interactions occur for $K \gg 1$. 
Carrying out the sums in Eq.~\eqref{result_distance_dependence} results in
\begin{equation}
    \gamma_S^n= c^{2n}\gamma_S^0 + (1-c^{2n})\gamma_A + \frac{2Ks^2}{cK-1}\left( \frac{c^2-c^{2n}}{s^2}- \frac{c^{n-1}K^{1-n}-1}{1-c^{-1}K} \right)\zeta.
     \label{exp_solution}
\end{equation}
In the limit $n\to \infty$ one finds the steady-state 
\begin{equation}
    \gamma_{S_{exp}}^{ss}=\gamma_A + \frac{2cK}{K -c}\zeta. 
    \label{result_exp_ss}
\end{equation}
This reduces to Eq.~\eqref{result_firstneighbor_ss} when $K \to \infty$. 
Since $K > 1$, the second term in~\eqref{result_exp_ss} never diverges. 
But if one has $K \gtrsim 1$ (long-ranged correlations), it may become very large.
The prefactor $2cK/(K-c)$ is shown in 
Fig.~\ref{fig:correlations_influence_ss} as a function of $\tau$.
It is always zero when $\tau = \pi/2$, and can  be negative if $\tau \in [\pi/2,3\pi/2]$.

In the two examples above, we only specify the dependence of the correlations with the distance $d$ between ancillas. 
However, $\zeta_d$ is still a $2\times 2$ matrix and we still have the flexibility of specifying its different entries. 
This opens up new and interesting possibilities. 
For instance, we may have a situation where $\gamma_A$ is a thermal state of the form 
\begin{equation}
    \gamma_A = (\mathcal{N}_A+1/2) \mathbb{I}_2,
    \label{CV_gamma_A_thermal}
\end{equation}
where $\mathcal{N}_A$ is a Bose-Einstein occupation. 
In the absence of correlations the system would thus reach a thermal state.
But when correlations are present, the steady-state~\eqref{result_firstneighbor_ss} may instead be a squeezed-state. 
This is an example in which the correlations between the ancillas actually produce squeezing in the system.

\subsection{\label{sec:gaussian_initial_states}Gaussian graph states for the ancillas}

Lastly, we  turn to the question of how one may  construct translationally invariant ancilla Gaussian initial states of the form~\eqref{CV_gamma_A}.
One possibility is through the 2-step collision model studied in~\cite{Seah2019}. 
Instead, here we focus in the case of Hamiltonian graph states, as we did for qubits in Sec.~\ref{sec:qubit}.
We have found it convenient to use the approach put forth in~\cite{Menicucci2011}, where the interaction Hamiltonian is taken to be a two-mode squeezing interaction
\begin{equation}
    H_{i,j} = \frac{i}{2} (a_i^\dagger a_j^\dagger - a_i a_j).
\end{equation}
The effect of the operator $\mathcal{V} = e^{-i k \sum_{i,j} G_{i,j} H_{i,j}}$, entering Eq.~\eqref{psi_A}, is to transform the quadrature operators according to 
\begin{equation}
    \mathcal{V}^\dagger q_i \mathcal{V} = \sum\limits_j M_{ij} q_j, 
    \qquad 
    \mathcal{V}^\dagger p_i \mathcal{V} = \sum\limits_j (M^{-1})_{ij} p_j, 
\end{equation}
where
\begin{equation}\label{MGk}
    M = e^{G k},
\end{equation}
is the matrix exponential of the adjacency matrix $G$. 

One may now directly use this to compute the ancilla CM $\gamma_{\bm A}$. 
We choose the state $|\phi\rangle$ in~\eqref{psi_A} as the vacuum.
As a consequence, $|\Psi_{\bm A}\rangle$ will be a multi-mode squeezed state. 
The expectation values over $|\Psi_{\bm A}\rangle$  yield
\begin{IEEEeqnarray}{rCl}
\label{CV_pos_corr}
    \frac{1}{2} \langle \{q_i, q_j\} \rangle &=& \frac{1}{2} (MM\trans)_{ij}, \\[0.2cm]
\label{CV_mom_corr}    
    \frac{1}{2} \langle \{p_i, p_j\} \rangle &=& 
    \frac{1}{2} [(M\trans M)^{-1}]_{ij}, \\[0.2cm]
    \langle q_i p_j \rangle &=& 0.
\end{IEEEeqnarray}
This allows us to construct $\gamma_{\bm A}$ for any graph state, irrespective of whether or not it is cyclic. 
The reduced ancilla CMs will be of the form 
\begin{equation}
\label{CV_gamma_A_i_gen}
    \gamma_{A_i} = \frac{1}{2}
    \begin{pmatrix}
    (MM\trans)_{ii} & 0 \\[0.2cm]
    0 & [(M\trans M)^{-1}]_{ii}
    \end{pmatrix},
\end{equation}
while the correlations will be
\begin{equation}
\label{CV_zeta_ij_gen}
    \zeta_{i,j} = \frac{1}{2} \begin{pmatrix}
    (MM\trans)_{i j} & 0 \\[0.2cm]
    0 & [(M\trans M)^{-1}]_{ij}
    \end{pmatrix}.
\end{equation}
Thus, to construct the CM, all we need to know is the matrix exponential of the graph adjacency matrix $M = e^{G k}$.

The object $e^{Gk}$ also appears in studies of complex networks.
In Ref.~\cite{Estrada2010}, for instance, it was shown that the diagonal entries $M_{ii}$ are a measure of the centrality of node $i$. That is, of how important this node is when serving as a hub for other connecting nodes.
Similarly, $M_{i,j}$ with $j\neq i$, is a measure of communicability between the two nodes. 
These two interpretations help to shed light on the interpretations of Eqs.~\eqref{CV_gamma_A_i_gen} and~\eqref{CV_zeta_ij_gen}. 
For instance, the local variances~\eqref{CV_gamma_A_i_gen}, which reflect the local uncertainty about each quadrature, are affected by both the graph centrality \emph{and} the communicability. 

Next we specialize this to cyclic graphs of the form~\eqref{G}. 
This provides yet another advantage to Gaussian states, since cyclic matrices can be diagonalized analytically as
\begin{equation}
G = \mathcal{O} \Lambda \mathcal{O}^\dagger,
\end{equation}
where the eigenvector matrix is given by the discrete Fourier transform
\begin{equation}
\mathcal{O}_{n,m} = \frac{e^{i 2\pi n m/N_A}}{\sqrt{N_A}}, \qquad n,m=0,\ldots,N_A-1
\end{equation}
and the eigenvalue matrix $\Lambda_{n,m} = \delta_{n,m} \lambda_m$ has 
\begin{equation}\label{CV_eig_vals}
    \lambda_m = 2\sum\limits_{\ell=1}^{N_A/2}  c_\ell \cos(2\pi \ell m/N_A),
\end{equation}
where we assumed $N_A$ was even for simplicity. 
The eigenvalues depend on the choice of coefficients $c_i$ in Eq.~\eqref{G}, but the eigenvectors do not. 

Using these results one readily finds that 
\begin{equation}
    M_{j\ell} = \frac{1}{N_A} \sum\limits_{m=0}^{N_A-1} e^{i 2\pi (j-\ell) m/N_A + k \lambda_m},
\end{equation}
Plugging this in Eqs.~\eqref{CV_pos_corr} and~\eqref{CV_mom_corr} gives us the reduced ancilla CMs $\gamma_{A_i} \equiv \gamma_A$, as well as the correlations $\zeta_{j,j+d} \equiv \zeta_d$.
First, 
\begin{equation}
\label{CV_graph_gammaA}
    \gamma_A = \frac{1}{2N_A}\begin{pmatrix}
        \sum\limits_{m=0}^{N_A-1} e^{ 2k \lambda_m} & 0 \\[0.2cm]
        0 & \sum\limits_{m=0}^{N_A-1} e^{-2k \lambda_m}
    \end{pmatrix}
\end{equation}
As can be seen, these are squeezed thermal states, whose properties are determined by the eigenvalues $\lambda_m$ (and hence the coefficients $c_i$ in~\eqref{G}). 

Next, the correlation matrices have the form 
\begin{equation}
\label{CV_graph_corr_mat}
    \zeta_d = \begin{pmatrix}
    \zeta_d^{(q)} & 0 \\[0.2cm]
    0 & \zeta_d^{(p)}. 
    \end{pmatrix}
\end{equation}
where 
\begin{IEEEeqnarray}{rCl}
\label{CV_graph_corr_q}
\zeta_d^{(q)} &=& \langle q_j q_{j+d}\rangle = \frac{1}{2N_A} \sum\limits_{m = 0}^{N_A-1} e^{i2\pi  dm/N_A + 2 k \lambda_m}, \\[0.2cm]
\zeta_d^{(q)} &=& \langle p_j p_{j+d}\rangle = \frac{1}{2N_A} \sum\limits_{m = 0}^{N_A-1} e^{i2\pi d m/N_A - 2 k \lambda_m}.
\label{CV_graph_corr_p}
\end{IEEEeqnarray}
We now illustrate the physics behind these results with some simple graphs.

\paragraph{Example: nearest-neighbor graph.} We assume $c_1 = 1$ and $c_i = 0$ for $i> 1$. Then $\lambda_m = 2 \cos(2\pi m/N_A)$. 
The sums in Eq.~\eqref{CV_graph_gammaA} can be computed exactly when $N_A$ is large by converting them to integrals
\begin{equation}
    \frac{1}{N_A} \sum\limits_{m=0}^{N_A-1} e^{\pm 4k \cos(2\pi m/N_A)} \to \frac{1}{\pi} \int\limits_0^\pi dx~e^{\pm 4 k \cos (x)} = I_0(4k),
\end{equation}
where $I_0$ is the modified Bessel function of the first kind.
The variances of $q$ and $p$ turn out to be the same, which is very particular of nearest-neighbors
The reduced state is thus actually a thermal state 
\begin{equation}
    \gamma_A = \frac{I_0(4k)}{2}~\mathbb{I}_2,
\end{equation}
with a Bose-Einstein occupation [Eq.~\eqref{CV_gamma_A_thermal}] determined by the value of $k$.

\begin{figure}
    \centering
    \includegraphics[width=0.45\textwidth]{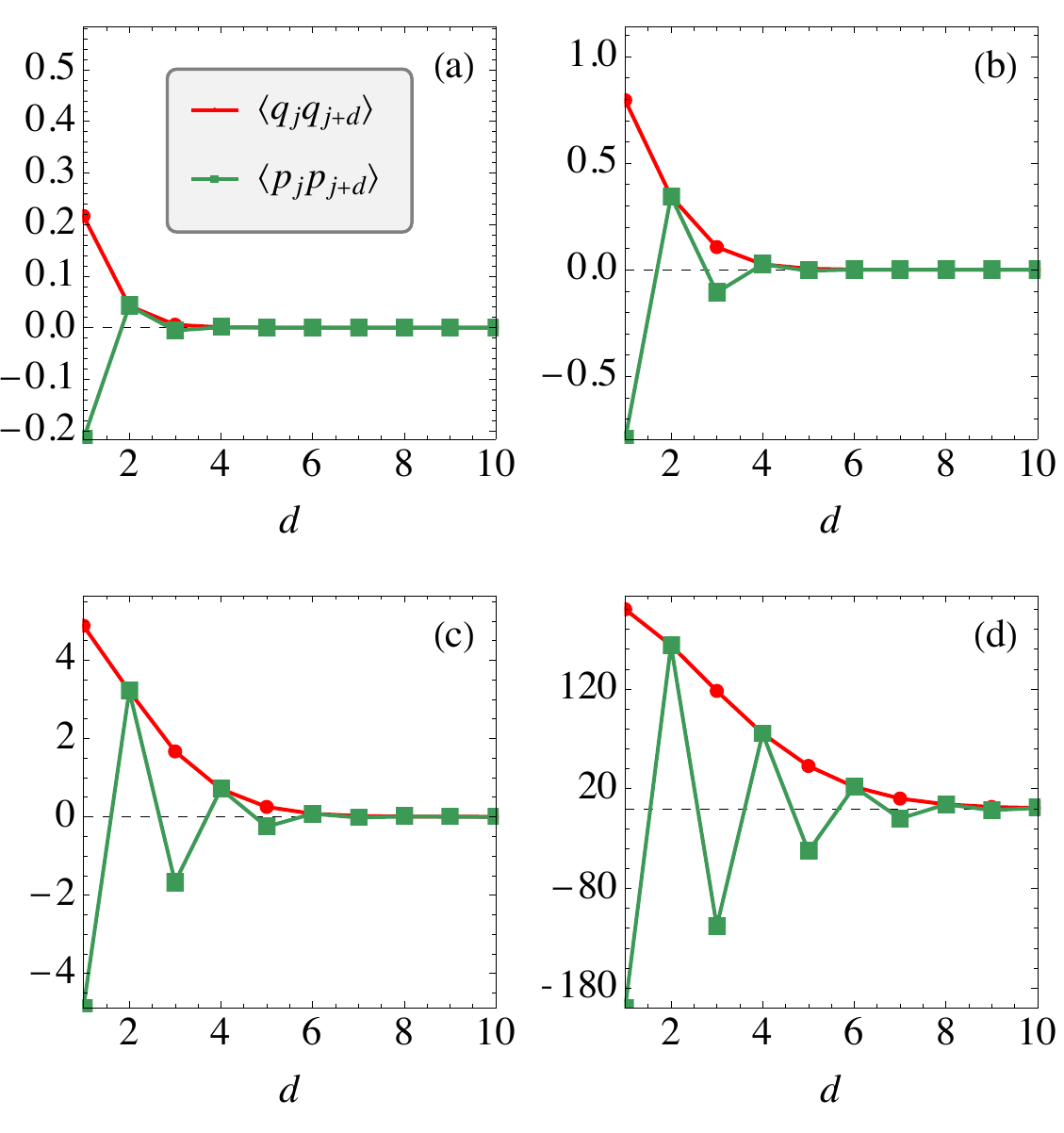}
    \caption{Correlation functions for the initial ancilla state, Eqs.~\eqref{CV_graph_corr_q} and~\eqref{CV_graph_corr_p}, for $k = 0.2, 0.5, 1.0, 2.0$.
    The position and momentum correlations actually have the same absolute values, but the sign of the latter oscillates with $d$.
    }
    \label{fig:CV_nn_corr}
\end{figure}

The correlations~\eqref{CV_graph_corr_q} and~\eqref{CV_graph_corr_p}, on the other hand, are plotted in Fig.~\ref{fig:CV_nn_corr}.
The figures were computed with $N_A = 100$, but the results concern only the first few neighbors and are thus independent of $N_A$, provided it is sufficiently large. 
As can be seen, the correlations decay with the distance $d$. 
Both  have the same magnitude (again, particular of nearest-neighbors), but the momentum correlations are oscillatory. 
The situation, of course, could be inverted by taking the interaction strength $k<0$.

\begin{figure}
    \centering
    \includegraphics[width=0.45\textwidth]{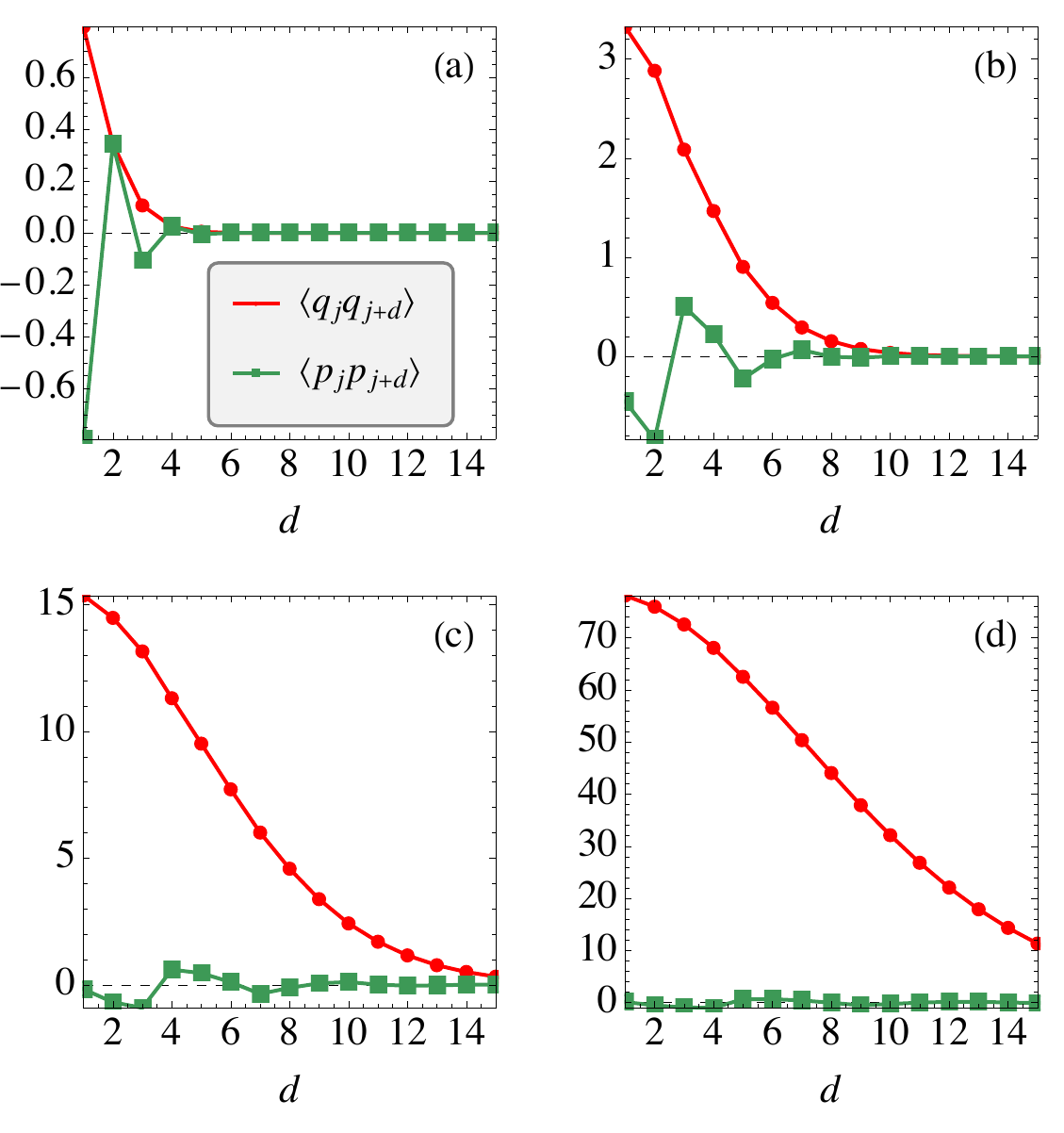}
    \caption{Similar to Fig.~\ref{fig:CV_nn_corr}, but varying the number of nearest-neighbors, from 1 to 4, with fixed $k = 0.7$. Image (a) is the same as Fig.~\ref{fig:CV_nn_corr}(a).
    }
    \label{fig:CV_nn_corr2}
\end{figure}

\paragraph{Example: $2^{\rm nd}$, $3^{\rm nd}$ and $4^{\rm nd}$ nearest-neighbors.} For comparison, we also analyze what happens if we choose graphs with second, third and fourth nearest-neighbors.
The results are shown in Fig.~\ref{fig:CV_nn_corr2}. 
As can be seen, while the momentum correlations (green squares) remain of order unity, the position correlations increase significantly with increasing number of neighbors. 

{\color{black}
\subsection{Perturbative treatment and scaling of correlations}

Finally, we use the results from the previous section to address the following question: \emph{in the case of weak correlations between the ancillas, is there any universal rule dictating by which degree the homogenization is broken?}
To make the question fair, we must compare $\gamma_S^n$ with the case in which the ancillas are uncorrelated, but otherwise have the same local state. 
From Eq.~\eqref{result_distance_dependence} we therefore see that the amount by which homogenization is broken is given by the last term
\begin{equation}\label{perturbation}
    \Delta \gamma_S^n = 2 s^2 \sum\limits_{m=1}^{n-1} c^{2m} \sum\limits_{d=1}^m c^{-d} \zeta_d. 
\end{equation}
The mismatch is thus linear in $\zeta_d$. 

Let us now consider ancillas prepared in a Hamiltonian graph state given by Eqs.~\eqref{CV_gamma_A_i_gen} and~\eqref{CV_zeta_ij_gen}. 
We assume that the overall strength $k$ of the correlations, in Eq.~\eqref{MGk}, is small and can thus be treated perturbatively. 
We can then expand 
\begin{equation}
    MM\trans = \mathbb{I} + 2 G k + 2 G^2 k^2 + \ldots.
\end{equation}
The adjacency matrix has zero diagonals.
Hence, to leading order in $k$, we may approximate the ancilla states as 
\begin{equation}
    \gamma_{A_i} \simeq \left[\frac{1}{2} + (G^2)_{ii}k^2 \right]\mathbb{I}_2, 
    \qquad 
    \zeta_{ij} \simeq G_{ij}k. 
\end{equation}
In light of Eq.~\eqref{perturbation}, we thus see that the correlations will contribute with a term of order $k$ to the breaking of homogenization. 
This scaling is universal and holds for any kind of graph state. 
Whether or not it holds for other types of correlated states (or beyond the continuous variable paradigm) is an open question, which we leave for future research.

}

\section{Discussions}

The goal of this paper was to show that  correlations among a system of ancillas can break homogenization. 
Locally it is as if the system is interacting with identical ancillas.
But globally, the correlations affect this dynamics and steer the system away from the homogenized fixed point. 
We studied this effect in two toy models. 
The first is a minimal qubit-qubit model, which is very similar to the original proposal for homogenization in Refs.~\cite{Ziman2002,Scarani2002}.
This model has the disadvantage that all calculations must be done numerically and, due to the exponential increase in Hilbert space dimension, we were only able to reach sizes of about $N_A = 16$ ancillas. 

In order to shed further light on the problem, we therefore switched to continuous-variable Gaussian dynamics. 
This allows for analytical results which neatly illustrate the non-trivial effects of correlations. 
Equations such as~\eqref{result_firstneighbor_ss} or~\eqref{result_exp_ss}, for instance, clearly show how the fixed point may differ dramatically from the reduced ancilla state $\gamma_A$. 


For concreteness, we have assumed that the initial ancilla state is translationally invariant. 
This introduces a nice interpretation where the system is interacting with elements which are locally identical. 
But, of course, it does not necessarily have to be this way. 
Several other types of interesting states may also be studied. 
One example would be to random graphs, so that the local states $\rho_{A_i}$ may fluctuate. 
In this case, it is expected that for partial swaps the system should eventually homogenize to their arithmetic mean. 
Notwithstanding, the overall effects of correlations should remain the same.

Another potentially interesting idea for a future study would be to analyze how the system affects the correlation profiles within the ancillas, after a series of collisions. 
This could be analyzed, for instance, using the framework put forth in~\cite{Girolami2017} for characterizing correlations in many-body states. 
Their framework allows one to construct quantifiers, called \emph{weaving}, which are specifically designed to understand how correlations change with system size $N_A$: 
Different classes of many-body states  scale with $N_A$ in different ways (algebraically, logarithmically, etc.). 
It would be particularly interesting to understand whether the collisions with the system can fundamentally alter the weaving. 
For qubit models, this may likely not be very illuminating since only small sizes can be reached. But for continuous models it should be possible. 

{\color{black}
Our formalism could in principle be implemented in current state-of-the-art quantum coherent platforms. 
The main difficulty is in the construction of the global ancilla state. 
Platforms such as trapped ions or nuclear spins could in principle be used for experiments with $\sim 10$ ancillas. 
Cavity quantum electrodynamics, with suitably prepared ancillas, is also a possibility. 
An experimentally more friendly approach would be to construct the correlations among the ancillas ``on the go.'' 
For instance, in the graph states just presented, the correlations generally fall quickly with the distance. 
Hence, in principle if one has access to recyclable ancillas, a suitably chosen periodic set of gates applied before the collisions could be used to always ensure that each ancilla is correlated with some of its neighbors. 

}

\begin{acknowledgements}

The authors would like to thank K. Modi for the interesting discussions concerning homogenization and thermodynamics. 
G.T.L. thanks the financial support of the S\~ao Paulo funding agency (FAPESP).
N.E.C. acknowledges the financial support from the Brazilian funding agency CNPq.

\end{acknowledgements}

\bibliography{library}
\end{document}